\documentclass[aps,prl,10pt,twocolumn,nofootinbib,superscriptaddress,preprintnumbers]{revtex4}
\usepackage{slashed}
\usepackage{graphicx}
\usepackage{times}
\usepackage{epsfig}
\usepackage{amsmath}
\usepackage{amsfonts}
\usepackage{amssymb}
\usepackage{url}
\usepackage{hyperref}
\usepackage{subfigure}
\newcommand{\be}{\begin{equation}}
\newcommand{\ee}{\end{equation}}
\newcommand{\bea}{\begin{eqnarray}}
\newcommand{\eea}{\end{eqnarray}}
\newcommand{\bref}[1]{(\ref{#1})}
\begin{document}
\pagestyle{plain}
\title{Bosonic Seesaw in the Unparticle Physics}
\author{Tatsuru Kikuchi}
\email{tatsuru@post.kek.jp}
\affiliation{Theory Division, KEK,
1-1 Oho, Tsukuba, 305-0801, Japan.}
\date{\today}
\begin{abstract}
Recently, conceptually new physics beyond the Standard Model 
 has been proposed by Georgi, where a new physics sector 
 becomes conformal and provides ``unparticle'' 
 which couples to the Standard Model sector through higher 
 dimensional operators in low energy effective theory. 
Among several possibilities, we focus on operators 
 involving the (scalar) unparticle, Higgs and the gauge bosons. 
Once the Higgs develops the vacuum expectation value (VEV), 
 the conformal symmetry is broken and as a result, 
 the mixing between the unparticle and the Higgs boson emerges.  
In this paper, we consider a natural realization of bosonic seesaw 
in the context of unparticle physics.
In this framework, the negative mass squared or the electroweak symmetry breaking
vacuum is achieved as a result of mass matrix diagonalization.
In the diagonalization process, it is important to have zero value in the $(1,1)$-element
of the mass matrix. 
In fact, the conformal invariance in the hidden sector can actually assure the zero
of  that element. So, the bosonic seesaw mechanism for the electroweak symmetry breaking 
can naturally be understood in the framework of unparticle physics.
\end{abstract}
\maketitle
\section{Introduction}
In spite of the success of the Standard Model (SM) 
 in describing all the existing experimental data, 
 the Higgs boson, which is responsible for the electroweak 
 symmetry breaking, has not yet been directly observed, 
 and is one of the main targets 
 at the CERN Large Hadron Collider (LHC). 
At the LHC, the main production process of Higgs boson is 
 through gluon fusion, and if Higgs boson is light, say 
 $m_h \lesssim 150$ GeV, the primary discovery mode 
 is through its decay into two photons. 
In the SM, these processes occur only at the loop level 
 and Higgs boson couples with gluons and photons very weakly. 

A certain class of new physics models includes 
 a scalar field which is singlet under the SM gauge group. 
In general, such a scalar field can mix with the Higgs boson 
 and also can directly couple with gluons and photons 
 through higher dimensional operators 
 with a cutoff in effective low energy theory.   
Even if the cutoff scale is very high, say, 100-1000 TeV, 
 the couplings with gluons and photons can be comparable 
 to or even larger than those of the Higgs boson 
 induced only at the loop level in the SM. 
This fact implies that if such a new physics exists, 
 it potentially has an impact on Higgs boson phenomenology 
 at the LHC. 
In other words, such a new physics may be observed together 
 with the discovery of Higgs boson. 

As one of such models, in this letter, 
 we investigate a new physics recently proposed 
 by Georgi \cite{Georgi:2007ek}, which is described in terms 
 of "unparticle" provided by a hidden conformal sector 
 in low energy effective theory. 
A concrete example of unparticle staff was proposed by Banks-Zaks \cite{Banks:1981nn} 
many years ago, where providing a suitable number of massless fermions, 
 theory reaches a non-trivial infrared fixed points 
 and a conformal theory can be realized at a low energy. 
Various phenomenological considerations on the unparticle physics 
 have been developed in the literature \cite{U-propagator, U-pheno}. 
It has been found that inclusion of the mass term for the unparticle plays an important role especially 
in studying about the Higgs-unparticle systems \cite{U-Higgs}, indeed we have studied the unparticle
physics focusing on the Higgs phenomenology including the effects of the conformal symmetry breaking \cite{Kikuchi}, 
and there are some other studies on the Higgs phenomenology in the literature of the unparticle physics \cite{U-Higgs2}.
Inclusion of such effects of the conformal symmetry breaking or the infrared (IR) cutoff is also considered in the literature of 
hadron collider physics \cite{Rizzo}, and in the model of colored unparticles \cite{U-color}.
There has also been studied on the astrophysical and cosmological applications of the unparticle physics \cite{U-astro},
especially, we have proposed a possibility for the unparticle dark matter scenario \cite{UDM}.
And there are some studies on the more formal aspects of the unparticle physics \cite{U-formal}
and its effects to the Hawking radiation \cite{Dai:2008qn}.

Now we begin with a review of the basic structure of the unparticle physics. 
First, we introduce a coupling between the new physics operator 
 ($\cal{O}_{\rm UV}$) with dimension $d_{\rm UV}$ 
 and the Standard Model one (${\cal O}_{\rm SM}$) with dimension $n$, 
\bea
 {\cal L} = \frac{c_n}{M^{d_{\rm UV}+n-4}} 
     \cal{O}_{\rm UV} {\cal O}_{\rm SM} ,  
\eea
where $c_n$ is a dimension-less constant, and $M$ is the energy scale 
 characterizing the new physics. 
This new physics sector is assumed to become conformal 
 at a energy $\Lambda_{\cal U}$, and 
 the operator $\cal{O}_{\rm UV}$ flows to the unparticle operator 
 ${\cal U}$ with dimension $d_{\cal U}$. 
In low energy effective theory, we have the operator of the form, 
\bea
{\cal L}=c_n 
 \frac{\Lambda_{\cal U}^{d_{\rm UV} - d_{\cal U}}}{M^{d_{\rm UV}+n-4}}   
 {\cal U} {\cal O}_{\rm SM} 
\equiv 
  \frac{1}{\Lambda^{d_{\cal U}+ n -4}}  {\cal U} {\cal O}_{\rm SM},  
\eea 
where the dimension of the unparticle ${\cal U}$ have been 
 matched by $\Lambda_{\cal U}$ which is induced 
 the dimensional transmutation, 
 and $\Lambda$ is the (effective) cutoff scale of 
 low energy effective theory. 
In this paper, we consider only the scalar unparticle. 

It was found in Ref.~\cite{Georgi:2007ek}
 that, by exploiting scale invariance of the unparticle, 
 the phase space for an unparticle operator 
 with the scale dimension $d_{\cal U}$ and momentum $p$ 
 is the same as the phase space for 
 $d_{\cal U}$ invisible massless particles, 
\begin{eqnarray}
d \Phi_{\cal U}(p) = 
 A_{d_{\cal U}} \theta(p^0) \theta(p^2)(p^2)^{d_{\cal U}-2} 
 \frac{d^4p}{(2\pi)^4} \,,
\label{Phi}
\end{eqnarray}
where
\begin{eqnarray}
A_{d_{\cal U}} = \frac{16 \pi^{\frac{5}{2}}}{(2\pi)^{2 d_{\cal U}}}
\frac{\Gamma(d_{\cal U}+\frac{1}{2})}{\Gamma(d_{\cal U}-1) 
\Gamma(2 d_{\cal U})}.
\label{A}
\end{eqnarray}
%
Also, based on the argument on the scale invariance, 
 the (scalar) propagator for the unparticle was suggested to be \cite{U-propagator}
\begin{eqnarray}
 \frac{A_{d_{\cal U}}}{2\sin(\pi d_{\cal U})}
 \frac{i}{(p^2)^{2-d_{\cal U}}} 
 e^{-i (d_{\cal U}-2) \pi}  \,.
\label{propagator}
\end{eqnarray}
%
Because of its unusual mass dimension, 
 unparticle wave function behaves as 
 $\sim  (p^2)^{(d_{\cal U}-1)/2}$ (in the case of scalar unparticle). 
 
\section{Unparticle and the Higgs sector}
First, we begin with a brief review of our previous work on
the Higgs phenomenology in the unparticle physics \cite{Kikuchi}.
Among several possibilities, we will focus on the operators 
 which include the unparticle and the Higgs sector, 
\be 
 {\cal L} = \frac{1}{\Lambda^{d_{\cal U}+ n - 4}} 
 {\cal U} {\cal O}_{\rm SM}(H^\dagger H)  
+\frac{1}{\Lambda^{2 d_{\cal U} +n -4}} 
{\cal U}^2 {\cal O}_{\rm SM}(H^\dagger H)  \;,
\ee
where $H$ is the Standard Model Higgs doublet and  
 ${\cal O}_{\rm SM}(H^\dagger H)$ is the Standard Model 
 operator as a function of the gauge invariant 
 bi-linear of the Higgs doublet. 
Once the Higgs doublet develops the VEV, 
 the tadpole term for the unparticle operator is induced,
\bea 
{\cal L}_{\slashed{\cal U}} =
 \Lambda_{\slashed{\cal U}}^{4-d_{{\cal U}}} {\cal U},  
\label{tadpole} 
\eea 
 and the conformal symmetry in the new physics sector is broken 
 \cite{U-Higgs}. 
Here, 
   $ \Lambda_{\slashed{\cal U}}^{4-d_{{\cal U}}}= 
   \langle {\cal O}_{\rm SM} \rangle/ \Lambda^{d_{\cal U}+n-4}$ 
 is the conformal symmetry breaking scale. 
At the same time, we have the interaction terms 
 between the unparticle and the physical Standard Model Higgs boson 
 ($h$) such as (up to ${\cal O}(1)$ coefficients) 
\bea 
  {\cal L}_{{\cal U}-{\rm Higgs}} 
  &=& \frac{\Lambda_{\slashed {\cal U}}^{4-d_{{\cal U}}}}{v} 
   {\cal U} h 
  + \frac{\Lambda_{\slashed {\cal U}}^{4-d_{{\cal U}}}}{v^2} 
    {\cal U} h^2
    \nonumber\\
  &+& \frac{\Lambda_{\slashed {\cal U}}^{4-2d_{{\cal U}}}}{v} 
   {\cal U}^2 h 
  + \frac{\Lambda_{\slashed {\cal U}}^{4-2 d_{{\cal U}}}}{v^2} 
    {\cal U}^2 h^2 + \cdots \;,   
 \label{mixing}
\eea 
where $v=246$ GeV is the Higgs VEV. 
In order not to cause a drastic change or instability 
 in the Higgs potential, 
 the scale of the conformal symmetry breaking  
 should naturally be smaller than the Higgs VEV, 
 $\Lambda_{\slashed {\cal U}} \lesssim v$. 
 When we define the `mass' of the unparticle as a coefficient of the second derivative
 of the Lagrangian with respect to the unparticle, ${\cal U}$,
then the mass of the unparticle can be obtained in the following form, 
$m_{\cal U}^{2-d_{{\cal U}}} = \Lambda_{\slashed {\cal U}}^{2-d_{{\cal U}}}$.

As operators between the unparticle and the Standard Model sector, 
 we consider 
\bea
{\cal L}_{\cal U} = 
 -\frac{\lambda_g}{4}  \frac{\cal U} {\Lambda^{d_{\cal U}}}
      G^A_{\mu \nu} G^{A \mu \nu} 
 -\frac{\lambda_\gamma}{4} \frac{\cal U}{\Lambda^{d_{\cal U}}} 
      F_{\mu \nu} F^{\mu \nu},  
\label{Unp-gauge} 
\eea
where we took into account of the two possible relative signs 
 of the coefficients, 
 $\lambda_g = \pm 1$ and $\lambda_\gamma= \pm 1$. 
We will see that these operators are the most important ones 
 relevant to the Higgs phenomenology.

Now let us focus on effective couplings 
 between the Higgs boson and the gauge bosons 
 (gluons and photons) of the form, 
\bea 
 {\cal L}_{\rm Higgs-gauge} = 
  \frac{1}{v} C_{gg} \;  h G^A_{\mu \nu} G^{A \mu \nu}  
+ \frac{1}{v} C_{\gamma \gamma} \; h F_{\mu \nu} F^{\mu \nu}.   
 \label{Higgs-gauge}
 \nonumber\\
\eea
As is well-known, in the Standard Model, these operators are induced 
 through loop corrections in which fermions and W-boson are running 
 \cite{HHG}. 
For the coupling between the Higgs boson and gluons, 
 the contribution from top quark loop dominates and is 
 described as
\bea 
 C_{gg}^{\rm SM} = \frac{\alpha_s}{16 \pi} F_{1/2}(\tau_t), 
 \label{SM-glue}
\eea
where $\alpha_s$ is the QCD coupling, 
 and $\tau_t = 4 m_t^2/m_h^2$ with the top quark mass $m_t$ 
 and the Higgs boson mass $m_h$.  
For the coupling between the Higgs boson and photons, 
 there are two dominant contributions from 
 loop corrections though top quark and W-boson, 
\bea 
 C_{\gamma \gamma}^{\rm SM} = 
  \frac{\alpha} {8 \pi} 
  \left( 
   \frac{4}{3} F_{1/2}(\tau_t) + F_{1}(\tau_W) 
  \right) ,     
\label{SM-gamma}
\eea 
where $\tau_W = 4 M_W^2/m_h^2$ with the W-boson mass $M_W$. 
In these expressions, the structure functions are defined as 
\begin{eqnarray}
F_{1/2}(\tau) &=& 
  2 \tau \left[ 1+ \left( 1 - \tau  \right) f(\tau) \right] , 
 \nonumber\\
F_{1}(\tau) &=& 
 -\left[2 + 3 \tau + 3 \left( 2 - \tau  \right) f(\tau) \right] 
\end{eqnarray}
with 
\begin{eqnarray}
f(\tau)  =  
   \left\{ 
    \begin{array}{cc}
       \left[ \sin^{-1}\left( 1/\sqrt{\tau}\right)\right]^2 &
               ({\rm for}~\tau\ge 1), \\
      -\frac{1}{4} 
    \left[ \ln \left(
       \frac{1+\sqrt{1-\tau}}{1-\sqrt{1-\tau}} \right)-i \pi \right]^2 &
               ({\rm for}~\tau < 1) . 
                      \end{array}  \right.   
\nonumber 
\end{eqnarray}
Note that even though the effective couplings are loop suppressed 
 in the Standard Model, they are the most important ones 
 for the Higgs boson search at the LHC and ILC. 
In the wide range of the Higgs boson mass $m_h < 1$ TeV, 
 the dominant Higgs boson production process at the LHC 
 is the gluon fusion channel though the first term 
 in Eq.~(\ref{Higgs-gauge}). 
If the Higgs boson is light, $m_h < 2 M_W$, 
 the primary discovery mode of the Higgs boson is 
 on its decay into two photons, 
 in spite of this branching ratio is ${\cal O}(10^{-3})$ at most. 
Therefore, a new physics will have a great impact on 
 the Higgs phenomenology at LHC and ILC. 
 if it can provide sizable contributions to 
 the effective couplings in Eq.~(\ref{Higgs-gauge}). 
Furthermore, the fact that the Standard Model contributions 
 are loop-suppressed implies that it is relatively easier 
 to obtain sizable (or sometimes big) effects from new physics. 

 Now we consider new contributions to the Higgs effective couplings 
 induced through the mixing between the unparticles 
 and the Higgs boson (the first term in Eq.~(\ref{mixing})) 
 and Eq.~(\ref{Unp-gauge}), 
 in other words, through the process  
 $h \to {\cal U} \to gg$  or $\gamma \gamma$. 
We can easily evaluate them 
  by using the vertex among the unparticle, 
 the Higgs boson and gauge bosons and the unparticle propagator as 
\bea
&& C^{\cal U}_{gg, \gamma \gamma} 
= \Lambda_{\slashed{\cal U}}^{4-d_{\cal U}} 
\left(
   \frac{A_{d_{\cal U}}}{2\sin(\pi d_{\cal U})} 
   \frac{e^{-i (d_{\cal U}-2) \pi}}{(m_h^2)^{2-d_{\cal U}}}  \right) 
\left( 
  \frac{\lambda_{g, \gamma}}{\Lambda^{d_{\cal U}}}
  \right)                                 
\nonumber\\
 &&=   \lambda_{g, \gamma} \; 
 \frac{A_{d_{\cal U}} e^{-i (d_{\cal U}-2) \pi}}{2\sin(\pi d_{\cal U})}
  \left(\frac{\Lambda_{\slashed {\cal U}}}{m_h} \right)^{4-d_{{\cal U}}}
  \left(\frac{m_h}{\Lambda} \right)^{d_{\cal U}} \;,  \quad \quad
\label{Unp-g}
\nonumber\\
\eea
where we replaced the momentum in the unparticle propagator 
 into the Higgs mass, $p^2 = m_h^2$. 
The unparticle contributions become smaller 
 as $m_h$ and $\Lambda$ ($\Lambda_{\slashed{\cal U}}$) 
 become larger (smaller) 
 for a fixed $1< d_{\cal U} <2$. 
Note that in the limit $d_{\cal U} \to 1$, 
 the unparticle behaves as a real scalar field and 
 the above formula reduces into the one obtained 
 through the mass-squared mixing $ \Lambda_{\slashed {\cal U}}^3/v$  
 between the real scalar and the Higgs boson.

Let us first show the partial decay width 
 of the Higgs boson into two gluons and two photons. 
Here we consider the ratio of the sum of the Standard Model 
 and unparticle contributions to the Standard Model one, 
\bea 
 R= \frac{ \Gamma^{{\rm SM}+{\cal U}}(h \to gg,\; \gamma \gamma)} 
      { \Gamma^{\rm SM}(h \to gg,\; \gamma \gamma)} 
 = \frac{
   |C^{\rm SM}_{gg,\; \gamma \gamma} +C^{\cal U}_{gg,\; \gamma \gamma}|^2 
   } 
   {|C^{\rm SM}_{gg,\; \gamma \gamma}|^2} . 
\nonumber\\
\eea
and we define the event number ratio ($r$), 
\bea 
 r &=& 
 \frac{
 \sigma^{{\rm SM}+{\cal U}}(gg \to h)  \times 
 {\rm BR}^{{\rm SM}+{\cal U}}(h \to \gamma \gamma) 
 } {
 \sigma^{\rm SM}(gg \to h)  \times 
 {\rm BR}^{\rm SM}(h \to \gamma \gamma) } \nonumber  \\
&=&
 \frac{ 
 |C^{\rm SM}_{gg}+C^{\cal U}_{gg}|^2 \times 
 |C^{\rm SM}_{\gamma \gamma} + C^{\cal U}_{\gamma \gamma}|^2 
 }{ 
 |C^{\rm SM}_{gg}|^2  \times |C^{\rm SM}_{\gamma \gamma}|^2 
 } .
\label{EventRatio}
\eea
Using Eqs.~(\ref{SM-glue}), ({\ref{SM-gamma}) and (\ref{Unp-g}) 
 we evaluate the ratio of the partial decay widths 
 as a function of $d_{\cal U}$. 
The numerical results in Ref. \cite{Kikuchi} show that,
even for $\Lambda={\cal O}$(1000 TeV), 
 we can see a sizable deviation of ${\cal O}(10 \%)$ 
 from the Standard Model one with $d_{\cal U} \sim 1$. 
Here, it is shown that the relative sign $\lambda_{g, \gamma}$ play 
 an important role in the interference between 
 the unparticle and the Standard Model contributions. 

As discussed before, once the Higgs doublet develops the VEV, 
 the conformal symmetry is broken in the new physics sector, 
 providing the tadpole term in Eq.~(\ref{tadpole}). 
Once such a tadpole term is induced, 
 the unparticle will subsequently develop the VEV
 \cite{U-Higgs, U-Higgs2} 
 whose order is naturally the same as the scale of 
 the conformal symmetry breaking, 
\bea
  \langle {\cal U} \rangle 
= \left( c \; \Lambda_{\slashed {\cal U}} \right)^{d_{{\cal U}}} . 
\eea 
Here we have introduced a numerical factor $c$, 
 which can be $c = {\cal O}(0.1) - {\cal O}(1)$, 
 depending on the naturalness criteria. 
Through this conformal symmetry breaking, 
 parameters in the model are severely constrained 
 by the current precision measurements. 
We follow the discussion in Ref.~\cite{U-Higgs}. 
 From Eq.~(\ref{Unp-gauge}), 
 the VEV of the unparticle leads to the modification of 
 the photon kinetic term, 
\bea
{\cal L} =
-\frac{1}{4}\left[ 
 1 \pm \frac{\langle {\cal U} \rangle}{\Lambda^{d_{{\cal U}}}}
\right] F_{\mu \nu} F^{\mu \nu} ,
\eea
 which can be interpreted as a threshold correction 
 in the gauge coupling evolution across the scale 
 $\langle {\cal U} \rangle^{1/d_{\cal U}}$. 
The evolution of the fine structure constant from 
 zero energy to the Z-pole is consistent 
 with the Standard Model prediction, 
 and the largest uncertainty arises 
 from the fine structure constant 
 measured at the Z-pole \cite{Yao:2006px}, 
\bea
 \widehat{\alpha}^{-1}(M_Z) &=&  127.918 \pm 0.019 . 
\nonumber 
\eea
This uncertainty (in the $\overline{\rm MS}$ scheme) 
 can be converted to the constraint,  
\bea
\epsilon=\frac{
\left<{\cal U} \right>}{\Lambda^{d_{{\cal U}}}}
 \lesssim 1.4 \times 10^{-4}. 
\label{gauge}
\eea
This provides a lower bound on the effective cutoff scale. 
For $d_{\cal U} \simeq 1$ and $\Lambda_{\slashed{\cal U}} \simeq v$ 
 we find 
\bea
\Lambda \gtrsim c \times 1000~{\rm TeV} ,  
 \nonumber 
\eea
This is a very severe constraint on the scale of new physics, 
 for example, $\Lambda \gtrsim 100$ TeV for $c \gtrsim 0.1$.

A similar bound can be obtained by the results on  
 Higgs boson search through two photon decay mode 
 at the Tevatron. 
With the integrated luminosity 1 fb$^{-1}$ and 
 the Higgs boson mass around $m_h=120$ GeV for example,  
 the ratio $r$ is constrained to be $r \lesssim 50$ \cite{Wells}. 
For $d_{\cal U} \simeq 1$, this leads to the bound, 
 $\Lambda \gtrsim 60$ TeV, which is, as far as we know, 
 the strongest constraint on the cutoff scale 
 by the current collider experiments. 
 
\section{Bosonic seesaw in the unparticle physics} 
Now we turn to the discussion of a realization of bosonic seesaw \cite{Calmet:2002rf, Kim:2005qb}
in the context of unparticle physics.
The basic idea of bosonic seesaw mechanism proposed in \cite{Calmet:2002rf, Kim:2005qb}
is to consider a bosonic analogy of the original seesaw mechanism \cite{seesaw}
in the neutrino physics.
In the case of neutrinos, the light neutrino mass eigenvalues are
obtained after diagonalizing the $2 \times 2$ matrix for one
generation.
\be
{\cal M}_\nu
=
\left(
\begin{array}{cc}
0 & m_D \\
m_D & M
\end{array}
\right)
\ee
where $m_D$ is the Dirac neutrino mass matrix and $M$ represents
the mass of the heavy right handed neutrino.

Assuming the heavy right handed Majorana mass scale $M \gg m_D$, 
the lightest mass eigenvalue of this mass matrix is given by
\be
m_\nu = - \frac{m_D^2}{M} \;.
\ee
There is no physical meaning of the minus sign In the case of neutrinos
since we can always absorb such a irrelevant phase factor by redefining 
the fields (rephasing).

However, in the case of scalar fields, such a phase factor possesses a physical meaning,
and we cannot erase it by redefinition of the fields.
This is a point for the bosonic seesaw works to make a mass squared of scalar fields
negative after the diagonalization.

The mass squared matrix of the unparticle-Higgs system is written by
\be
{\cal M}_h^2
=
\left(
\begin{array}{cc}
0 & \Lambda_{\slashed{\cal U}}^{4 - d_{\cal U}}/v \\
\Lambda_{\slashed{\cal U}}^{4 - d_{\cal U}}/v  & \lambda v^2 \\
\end{array}
\right) \;,
\label{Mh}
\ee
where $\lambda$ stands for the Higgs self coupling,
\be
{\cal L}_{\rm Higgs} = \mu_h^2 \, (H^\dag H) + \lambda \,(H^\dag H)^2
\ee
with $\lambda > 0$.

Assuming the conformal symmetry breaking scale in the hidden sector
is smaller than the weak scale, $\Lambda_{\slashed{\cal U}} \lesssim v$, 
then the eigenvalue of the mass matrix \bref{Mh} after diagonalization 
is given by
\be
\mu_h^2 = - \frac{\Lambda_{\slashed{\cal U}}^{8 - 2d_{\cal U}}}{\lambda v^4} \;.
\ee
Thereby, the negative mass squared $(\mu_h^2 < 0)$ or the electroweak symmetry breaking
vacuum is achieved as a result of the mass matrix diagonalization.
In this diagonalization process, it is important to have zero value in the $(1,1)$-element
of the mass matrix. 
In fact, the conformal invariance in the hidden sector can actually assure the zero
of  that element. So, the bosonic seesaw mechanism \cite{Calmet:2002rf, Kim:2005qb} for 
the electroweak symmetry breaking 
can naturally be understood in the framework of unparticle physics.

\section{Summary}
In conclusion, we have considered the unparticle physics 
 focusing on the Higgs phenomenology. 
Once the electroweak symmetry breaking occurs, 
 the conformal symmetry is also broken and 
 this breaking leads to the mixing between the unparticle 
 and the Higgs boson. 
Providing the operators among the unparticle and the gauge bosons 
 (gluons and photons),  
 the unparticle brings the sizable deviation into 
 effective couplings between the Higgs boson and the gauge bosons, 
 that can be measured at the LHC through the discovery of the Higgs boson.

In this paper, we specifically considered a realization of bosonic seesaw \cite{Calmet:2002rf, Kim:2005qb}
in the context of unparticle physics.
In this framework, the negative mass squared or the electroweak symmetry breaking
vacuum is achieved as a result of mass matrix diagonalization.
In the diagonalization process, it is important to have zero value in the $(1,1)$-element
of the mass matrix. 
In fact, the conformal invariance in the hidden sector can actually assure the zero
of  that element. So, the bosonic seesaw mechanism for the electroweak symmetry breaking 
can naturally be understood in the framework of unparticle physics.

\begin{center}
{\bf Acknowledgments}
\end{center}
We would like to thank N. Okada for his stimulating discussions.
The work of T.K. was supported by the Research
Fellowship of the Japan Society for the Promotion of Science (\#1911329).


\end{document}